\documentclass[aps,prl,showpacs,preprintnumbers,twocolumn]{revtex4-2}
\usepackage{amsmath,amssymb}
\usepackage{bm}
\usepackage{tipa}
\usepackage{upgreek}
\usepackage[english]{babel}
\babeladjust{autoload.bcp47 = on}

\usepackage{comment}
\usepackage{mathrsfs}
\usepackage{mathrsfs}
\usepackage{graphicx}
\usepackage{braket}
\usepackage{mathbbol}
\usepackage{booktabs}
\usepackage{enumitem}
\usepackage{amssymb}
\usepackage{enumitem}
\usepackage[normalem]{ulem}
\usepackage{arydshln}
\usepackage{color}
\usepackage[table,xcdraw]{xcolor}
\usepackage[colorlinks,bookmarks=true,citecolor=blue,linkcolor=red,urlcolor=blue]{hyperref}
\usepackage{cleveref}

\makeatletter
\let\old@makecaption=\@makecaption
\usepackage{subcaption}
\let\@makecaption=\old@makecaption
\makeatother

\usepackage{relsize}

\usepackage[skins]{tcolorbox}

\newtcolorbox{myframe}[2][]{%
  enhanced,colback=white,colframe=black,coltitle=black,
  sharp corners,boxrule=0.6pt,
  attach boxed title to top left={yshift=-0.3\baselineskip-0.4pt,xshift=2mm},
  boxed title style={tile,size=minimal,left=0.5mm,right=0.5mm,
    colback=white,before upper=\strut},
  title=#2,#1
}

\begin{document}

\title{Is the fractional Chern insulator–superconductor transition in twisted MoTe$_2$ direct?}

\author{Tevž Lotrič and Steven H. Simon}
\affiliation{Rudolf Peierls Centre for Theoretical Physics, Parks Road, Oxford, OX1 3PU, UK}

\begin{abstract}
The observation of superconducting behaviour near a fractional Chern insulator in twisted MoTe$_2$ has sparked excitement about a possible direct ``anyon superconducting'' transition between a topologically ordered and a symmetry-broken phase. We carefully examine the experimentally measured transport data in the relevant region of the low-temperature tMoTe$_2$ phase diagram and propose a scenario where the nearby reentrant integer quantum Hall effect (RIQHE) acts as an intervening phase between the two. While the RIQHE is most prominent at large displacement fields, we propose that it extends to smaller fields at very low temperatures. Based on this picture, we model the sample as a patchwork of the three phases and show that such a model can explain the main qualitative features of the transport phase diagram, including its dependence on temperature and displacement field. The model predicts that at lower electronic temperatures the apparent FCI–SC transition at zero displacement field should split into two transitions separated by a narrow RIQHE region.
\end{abstract}

\maketitle
\textit{Introduction---}
{The low-temperature phase diagram of twisted MoTe$_2$ displays a rich set of physical behaviours, the most exotic being the fractional Chern insulators~\cite{CaoReview} (FCIs, also known as fractional quantum anomalous Hall (FQAH) states). These are zero-field analogues of the celebrated fractional quantum Hall states, displaying the same quantized Hall response and charge fractionalization, generated by a spontaneous breaking of time-reversal symmetry. 
Twisted MoTe$_2$ therefore allows us to study the interplay between topological order and more conventional symmetry-breaking order in some detail. In particular, a superconducting state has been found near FCIs in Ref.~\cite{xu2026signaturesunconventionalsuperconductivitynear}. This triggered an influx of theoretical work asking about how these states may be connected, and exploring the possibly very exotic transition between the two. Some predictions are related to older theories of anyon superconductivity \cite{Chen:1989xs,WEN1990135,PhysRevB.103.035124e,kitazawa_exactness_1990}, but other mechanisms of how an FCI turns superconducting upon tuning the filling fraction have been proposed \cite{senthil_doping_2024,senthil_doping_2025,shi_anyon_2025,shi_non-abelian_2025,senthil2026fractionalizedmetalsdopedanyons,wang2025chiralsuperconductivitynearfractional,fan_hidden_2026,wang_topological_2026,wen_topological_2025,lotric2026phasesitinerantanyonslaughlins,guerci_fractionalization_2025,guerci_topological_2026,Nosov_2026,soejima_anyon_2024,chen_topological_2025,yan_anyon_2025,han_anyon_2025,Ma_2020,pichlerMicroscopicMechanismAnyon2025}, see \cite{mehta2026coloringanyonsuperconductivity,shi2026superconductivitynonfermiliquidmetals} for recent overviews. In particular, recent works have studied disorder-driven transitions and possible intermediate phases \cite{shi_anyon_2025}, as well as a unified description of the neighbouring superconducting and RIQHE states \cite{Nosov_2026}.

Instead of studying detailed microscopic mechanisms, we take a step back and carefully re-analyse some of the data in Ref.~\cite{xu2026signaturesunconventionalsuperconductivitynear}. Both the FCI and SC phases have vanishing longitudinal resistivity, $\rho_{xx}=0$, but as we move between them, a peak with $\rho_{xx}\lesssim h/e^2$ develops. By a careful analysis of the shape of this peak, we propose a different picture of the transition. We find that this peak is well explained by an inhomogeneous sample, where local patches realise either the FCI, SC or a third, re-entrant integer quantum Hall (RIQHE) phase, which is also nearby in the phase diagram. We conclude that, at sufficiently low temperatures, the system would display an intermediate RIQHE phase between the FCI and SC, meaning that the observed transport may not realise the proposed direct FCI-to-SC critical point. Both the RIQHE and SC phases have been interpreted as possibly emerging from the physics of itinerant anyons \cite{Nosov_2026}, however, so it remains possible that this region of the phase diagram is well-described by anyon physics.
}{ We note that some qualitatively similar transitions have been observed in rhombohedral graphene and have also been described in a multiphase picture with an effective medium theory~\cite{hadjri_quantum_2026}.   As we will describe below, our picture goes beyond effective medium, including spatial structure to properly describe the physics.}

\begin{figure}
    \centering
    \includegraphics[width=0.99\linewidth]{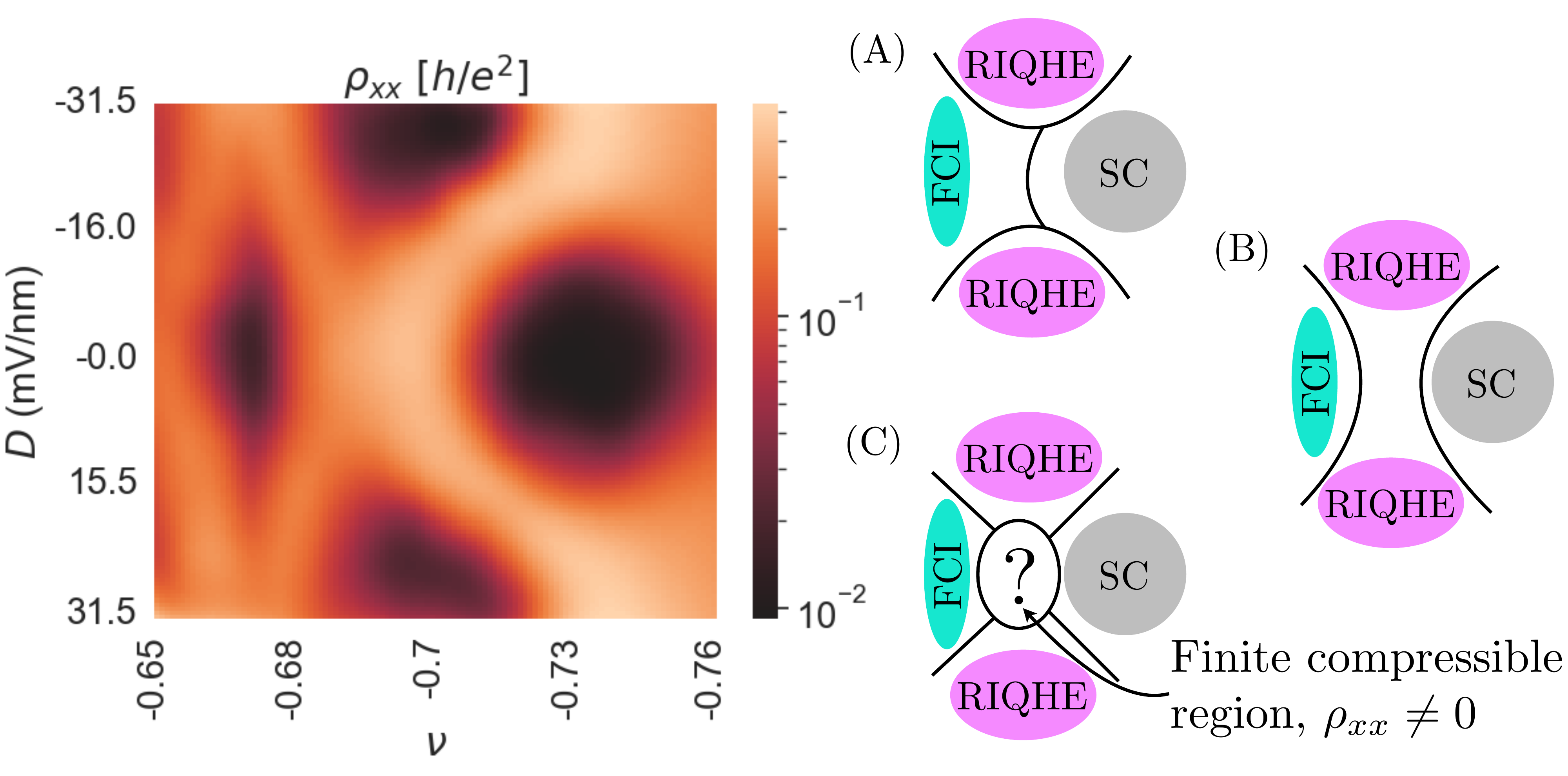}
    \caption{Left: Replotted $\rho_{xx}(\nu,D)$ from Ref.~\cite{xu2026signaturesunconventionalsuperconductivitynear}, showing four minima, associated with the fractional Chern insulator (FCI), reentrant integer quantum Hall (RIQHE), and superconducting (SC) phases at the lowest temperature achieved in experiment. The phases are separated by peaks in $\rho_{xx}$. Right: As temperature approaches zero, the resistive peaks are expected to become narrower. We sketch three possible natural topologies of the zero-temperature quantum phase diagram. In the present work, we argue that scenario (B), where the RIQHE is always an intervening phase between the FCI and SC, naturally explains the observed transport signatures. }
    \label{fig:phase_diagram_shapes}
\end{figure}

We focus on the transport phase diagram of tMoTe$_2$, measured in Ref.~\cite{xu2026signaturesunconventionalsuperconductivitynear} as a function of filling fraction $\nu$ and displacement field $D$, in the region $-0.76<\nu<-0.65$ and $|D|<32~\text{mV/nm}$. This region, replotted in Fig.~\ref{fig:phase_diagram_shapes}, displays four distinct local minima in the longitudinal resistivity $\rho_{xx}$, separated by resistive peaks. Based on measurements of the (anomalous) Hall resistivity $\rho_{xy}$ around $D=0$, a minimum around $\nu=-2/3$ is interpreted as an FCI state and another minimum near $\nu=-0.73$ as a superconductor. Around $\nu=-0.70$, another pair of minima is seen at $D=\pm30~\text{mV/nm}$ and the measured $\rho_{xy}=h/e^2$ identifies them as re-entrant integer quantum Hall states. 

The presence of these four phases complicates the extrapolation of the phase diagram in Fig.~\ref{fig:phase_diagram_shapes} to the low-temperature limit, where the wide resistive regions are expected to turn into narrow phase separators. In particular, several topologies of the phase diagram are possible, as sketched in Fig.~\ref{fig:phase_diagram_shapes}: (A) we could have a direct FCI-SC transition, with the RIQHE phases separated, (B) the two RIQHE minima could be connected, leaving no direct FCI-SC transition; or (C) a compressible phase may appear with finite $\rho_{xx}$ in the zero-temperature limit.  We now show that scenario (B) can explain the detailed shape of the resistive peak and its temperature dependence. {Alternative interpretations of the resistive structure neighbouring the RIQHE have also been proposed \cite{Huang_2025}.}

The strategy is to assume that in the region of interest, the three phases are in close energetic competition (this is supported numerically in a toy model \cite{wang2025chiralsuperconductivitynearfractional}). Then, even with a relatively small degree of disorder (be it twist angle disorder or impurities), one expects different phases to be favoured in different spatial regions, leading to a picture of an inhomogeneous sample containing puddles of different phases near the transitions. A variety of theories have been proposed for how continuous transitions between the various phases may look in the clean limit, but the exact details, including whether the clean transitions are first-order or continuous, do not matter in this picture. Disorder can lead to a puddled sample in both cases. 

The transport data provide several indications in favour of scenario (B). At high temperatures, the FCI–SC resistive peak approximately follows the two-phase semicircle law~\cite{PhysRevB.50.2369}, while on cooling, it develops a feature near $\rho_{xy}=h/e^2$, coincident with the emergence of the nearby RIQHE state at finite $D$. We show below that both observations arise naturally if the RIQHE intervenes between the FCI and SC but has a lower characteristic energy scale.

\textit{Observed resistivity semicircle---}
\begin{figure}
    \centering
    \includegraphics[width=0.99\linewidth]{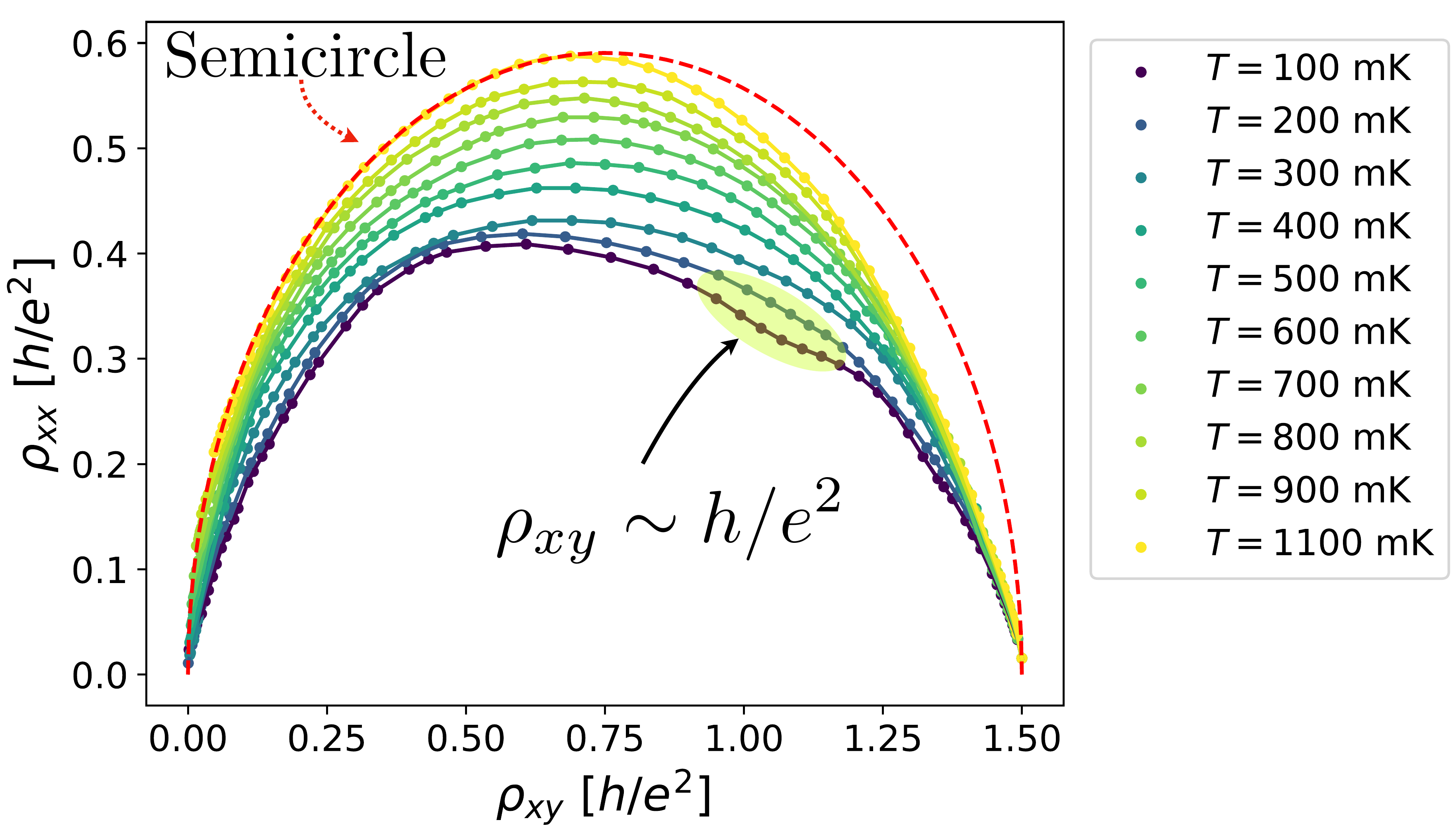} 
    \caption{The resistive peak in $\rho_{xx}$ as a function of $\rho_{xy}$ for various temperatures in the $D=0$ cut of the phase diagram. The higher temperatures show a dependence reminiscent of a scaled semicircle in the $(\rho_{xy},\rho_{xx})$ plane (red dashes), while at lower temperatures the peak in $\rho_{xx}$ is suppressed, and a knee in the data appears near $\rho_{xy}\sim h/e^2$. [Data from Ref.~\cite{xu2026signaturesunconventionalsuperconductivitynear}]}
    \label{fig:semicircle_thermal}
\end{figure}
We start by investigating the detailed shape of the resistive peak along the cut between the FCI and SC states in Fig.~\ref{fig:semicircle_thermal}. We take the $D=0$ cut for $\nu\in[-0.735,-0.666]$, showing the dependence of $\rho_{xx}$ on $\rho_{xy}$ at several temperatures. While one expects a phase transition to become a sharper feature as a function of external parameters in the zero-temperature case, with a narrow peak in $\rho_{xx}$ as a function of $\nu$, the shapes in Fig.~\ref{fig:semicircle_thermal} need not become sharply peaked. Instead, both $\rho_{xx}$ and $\rho_{xy}$ change sharply as functions of $\nu$, while their interdependence can still follow a smooth curve.

The first interesting observation is the shape of the $T=1100~\text{mK}$ curve, which is well described by a semicircle law $(S\rho_{xx})^2+(\rho_{xy}-3h/4e^2)^2=(3h/4e^2)^2$, the red dashed line in the figure \footnote{The semicircle law of Ref.~\cite{PhysRevB.50.2369} predicts $S=1$, while the best empirical fit in Fig.~\ref{fig:semicircle_thermal} is seen for $S\sim1.3$. In experiments, $\rho_{xx}=cR_{xx}$ for the measured longitudinal resistance $R_{xx}$ and a device-specific aspect ratio $c$. Part of the need for the re-scaling by $S$ might be explained by the use of the width-to-length ratio $c=W/L$ instead of a detailed simulation of current flows in the device. We note that \cite{hadjri_quantum_2026} recently observed semicircle behaviour for transitions out of FCIs in rhombohedral graphene, and similarly found the need to re-scale $\rho_{xx}$ by $\lesssim 30\%$.}. This form of the interdependence can be neatly explained by modelling the sample as a mixture of two phases undergoing a percolation transition in a regime where effects of quantum-coherent tunnelling over large distances can be ignored and the local conductivity satisfies the linear relation $\mathbf j(\mathbf r)=\sigma(\mathbf r)\mathbf E(\mathbf r)$, with a spatially dependent conductivity tensor $\sigma(\mathbf r)$ \cite{PhysRevB.50.2369,hilke_experimental_1998,hilke_semicircle_1999}. The shape of the high-temperature resistivity peak can thus be broadly explained as a percolation transition in a model of disorder-induced puddles of the FCI phase and a good conductor (as $T=1100~\text{mK}$ is above the superconductor's reported critical temperature). Such a model, however, cannot explain the decrease in the peak of $\rho_{xx}$ with temperature, and furthermore cannot produce the feature seen around $\rho_{xy}\sim h/e^2$ at the lowest temperatures. At $T>1~\text{K}$, the RIQHE resistivity minimum at $(\nu,D)\sim(-0.7,\pm 30~\text{mV/nm})$ shown in Fig.~\ref{fig:phase_diagram_shapes} is entirely absent. This suggests that the RIQHE has the lowest characteristic ordering energy scale of the three phases, preventing it from producing a distinct transport signature at this temperature. {Below the superconducting transition temperature $T_c$, coherent Cooper-pair transport can invalidate the local-conductivity approximation, particularly near the SC end of the curve. We therefore do not interpret low-temperature deviations there. Our argument instead relies on the feature near $\rho_{xy}=h/e^2$, which coincides with the Hall response of the nearby RIQHE.}

{Recently, Ref.~\cite{hadjri_quantum_2026} has similarly analysed shapes of curves in the $(\rho_{xy},\rho_{xx})$ plane for transitions between FCIs, IQHE-like states and Fermi-liquid states in rhombohedral graphene. While the microscopic system is different, they generally find good agreement of the results with the semicircle prediction, supporting a puddled picture of the sample near the transition. While Ref.~\cite{hadjri_quantum_2026} focused on known transitions, assessing their consistency with the semicircle, we pose an inverse question -- what can the shapes of the resistive peaks in the $(\rho_{xy},\rho_{xx})$ plane tell us about the nature of the mixtures that produce them?
}

\textit{The three-phase model---}
As temperature is decreased, the resistivity minimum associated with the RIQHE phase at $(\nu,D)\sim(-0.7,\pm 30~\text{mV/nm})$ becomes more pronounced. We now demonstrate how this decrease in $\rho_{xx}^\text{RI}$ is enough to explain the thermal dependence and low-temperature shape of Fig.~\ref{fig:semicircle_thermal}. We model spatially correlated quenched disorder in our sample with a scalar field $V(\mathbf r)=\mu+\xi(\mathbf r)$, where $\xi(\mathbf r)$ is a Gaussian random field with zero mean and correlation function $\langle\xi(\mathbf r)\xi(\mathbf r')\rangle = \exp(-|\mathbf r-\mathbf r'|^2/2l^2_\text{corr})$. We choose to work in length units where $l_\text{corr}=1$. Given a disorder configuration in the sample, each point $\mathbf r$ is identified as one of the three possible phases and assigned local resistivity values $\rho_{xx}(\mathbf r),~\rho_{xy}(\mathbf r)$ (in units where $h/e^2=1$) according to the rule
\begin{equation} \label{eq:phase_assignment_rule}
    [\rho_{xx}(\mathbf r),\rho_{xy}(\mathbf r)] \! =  \left\{ \begin{array}{lcll}   \![\epsilon~~~,\frac{3}{2}]~~~ & \text{if}& V(\mathbf r)<-w/2 &\text{(FCI)}\\
        \![\rho_{xx}^\text{RI},1]  &\text{if}& |V(\mathbf r)|\leq w/2 & \text{(RIQHE)}\\
        \![\epsilon~~~,0] &\text{if}& V(\mathbf r) > w/2 & \text{(SC)}. \end{array} \right.
\end{equation}
The parameter $w$ controls the relative prevalence of the RIQHE phase, while $\epsilon\ll1$ is the small resistivity assigned to the well-developed phases. The longitudinal resistivity of the RIQHE, $\rho_{xx}^\text{RI}$, is the crucial temperature-dependent control parameter we use to move between the different regimes in Fig.~\ref{fig:semicircle_thermal}.

The filling fractions at which the phases are most stable follow the sequence $\nu^\text{SC}<\nu^\text{RI}<\nu^\text{FCI}$, and Eq.~\ref{eq:phase_assignment_rule} is designed to respect this, with the RIQHE phase always appearing between domains of the FCI and SC. The role of the field $V(\mathbf r)$ is to generate random spatial configurations of the three phases; once the phase configuration is fixed, the field $V$ itself has no further bearing on the results. Giving this field a concrete physical interpretation would require a microscopic model of the disorder, which we do not pursue here. 

\begin{figure}
    \centering
    \includegraphics[width=0.98\linewidth]{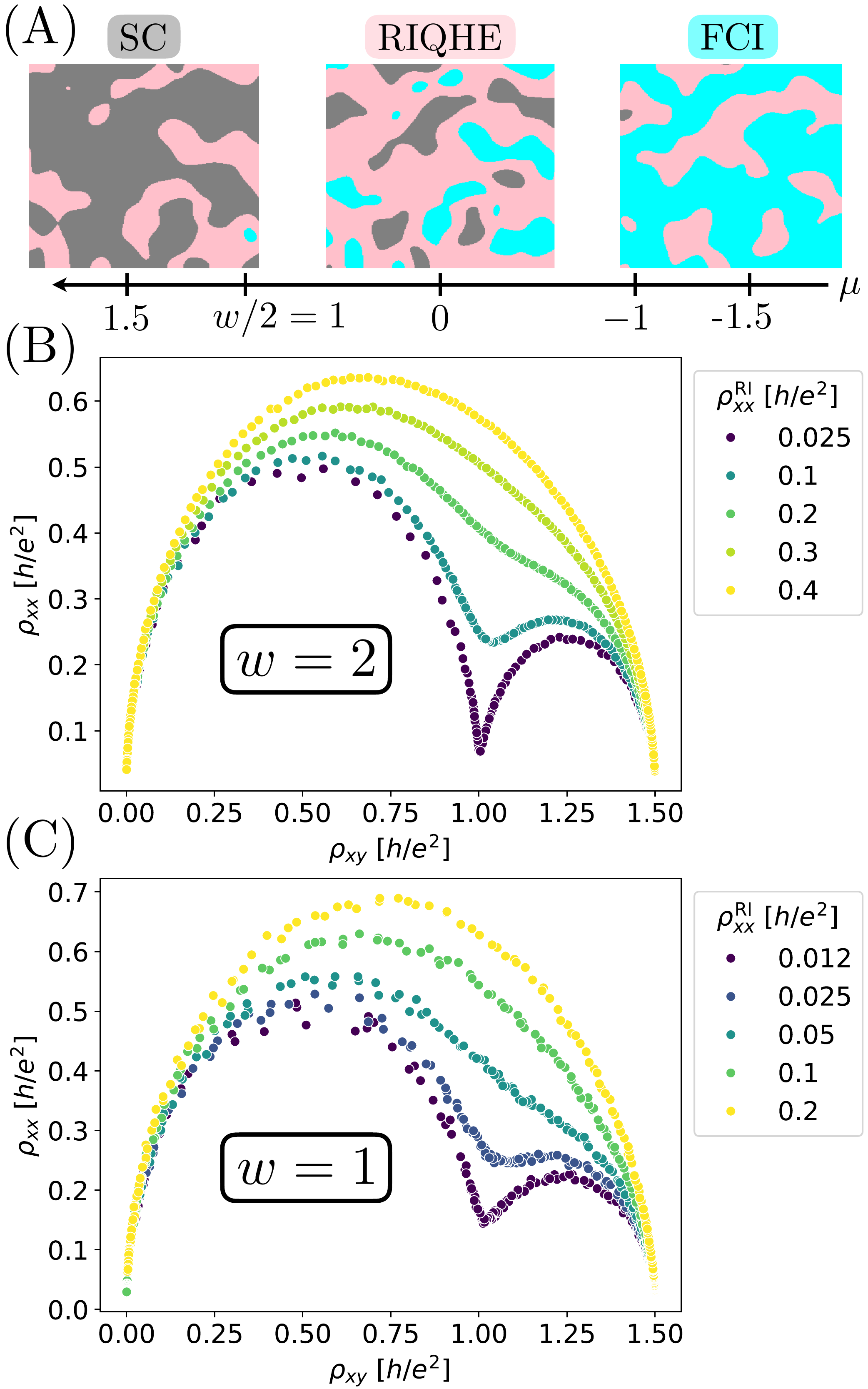}
    \caption{Theoretical evolution of the effective resistivity as the background potential $\mu$ is varied for different values of $\rho_{xx}^\text{RI}$. (A) Typical configurations evolve from being dominated by the FCI, to an intervening RIQHE phase, and finally into the SC phase. For all parameters, the RIQHE regions lie between the FCI and SC phases ($w=2$ for those plots). (B) The derived effective resistivity evolution. For a strong RIQHE ($\rho_{xx}^\text{RI}=\epsilon$), we see two semicircles, describing a pair of percolation transitions. For a weak RIQHE ($\rho_{xx}^\text{RI}=0.4h/e^2$), the resistivity is similar to that expected for an FCI-SC semicircle, and intermediate values interpolate between the extremes. Note that all values of $\rho_{xx}^\text{RI}$ use the same spatial configuration of the phases. (C) We repeat the analysis of (B), but with a smaller spatial fraction of the RIQHE phase, $w=1$. The resulting plots are qualitatively similar to (B), but for a given line shape, a smaller $\rho_{xx}^\text{RI}$ is needed. This is to be expected from narrower RIQHE regions. In both (B) and (C), we observe that the RIQHE minimum in resistivity, $\rho_{xx}^\text{RI-min}$, obtained when $\rho_{xy}=1$, is often significantly larger than the intrinsic resistivity of the RIQHE, $\rho_{xx}^\text{RI}$.  }   
    \label{fig:simulation_results}
\end{figure}

In this model, the physical transition is driven by varying the mean $\mu$ of the background field -- when $\mu\lesssim- w$, most of the sample is in the FCI phase, and we expect to measure $\rho_{xx}\sim0,\rho_{xy}\sim3/2$. At $\mu=0$, the RIQHE strip surrounds the percolating level set separating FCI and SC regions and therefore itself generally percolates for $w>0$. 
Finally, when $\mu\gtrsim w$, the sample is dominated by the SC patches, and we expect to see a superconducting response. Typical configurations for different values of $\mu$ are shown in Fig.~\ref{fig:simulation_results}(A). This can be visualised as a landscape $V(\mathbf r)$ filled with water to a fixed waterline.   Below the waterline is the SC phase, above the waterline is the FCI phase, and a region of width $w$ near the waterline is the RIQHE phase.
{
This spatial structure distinguishes our model from the three-component effective-medium approximation of Ref.~\cite{hadjri_quantum_2026}. Away from narrow transition regions, one component always percolates in our construction.
Consequently, when all component longitudinal resistivities vanish as $T\to0$, the macroscopic system reflects only the percolating phase rather than showing a mixture with some finite effective $\rho_{xx}$.
This allows us to identify the model directly with scenario (B) in Fig.~\ref{fig:phase_diagram_shapes}.
}

In Fig.~\ref{fig:simulation_results}(B), we choose $w=2$, set the residual FCI/SC resistivity to $\epsilon=0.025$, and sweep $\mu$, numerically computing the effective resistivity (details in the End Matter). Depending on how developed the RIQHE is, or equivalently on how small $\rho_{xx}^\text{RI}$ is, the behaviour is drastically different -- for a weak RIQHE ($\rho_{xx}^\text{RI}\sim0.4h/e^2$ --- a value in line with the high-temperature resistivity at the centre of the RIQHE phase at larger $|D|$), the resistivity behaves similarly to that expected for an FCI-SC semicircle. As we make the RIQHE more pronounced (decreasing $\rho_{xx}^\text{RI}$) while keeping the same distribution of phases, the height of the peak in $\rho_{xx}$ decreases. Around $\rho_{xx}^\text{RI}=0.2 h/e^2$, a feature develops near $\rho_{xy}\sim h/e^2$. This is qualitatively similar to the observed shapes in Fig.~\ref{fig:semicircle_thermal}. For even smaller $\rho_{xx}^\text{RI}$, the RIQHE-induced dip becomes significantly more pronounced.  We repeat the analysis with narrower RIQHE regions, $w=1$, in Fig.~\ref{fig:simulation_results}(C). The results are similar to those for $w=2$, but for a given shape of the line, i.e., for a given size of the RIQHE dip in the curve, a smaller value of $\rho_{xx}^\text{RI}$ is required. Still, the curve interpolates from a double transition as $\rho_{xx}^\text{RI}\to0$ to a broad semicircle-like shape when $\rho_{xx}^\text{RI}$ is large. {A large $\rho_{xx}^\text{RI}$ essentially broadens the two transitions until they resemble a single, wide peak in $\rho_{xx}$.}

\begin{figure}
    \centering
    \includegraphics[width=0.95\linewidth]{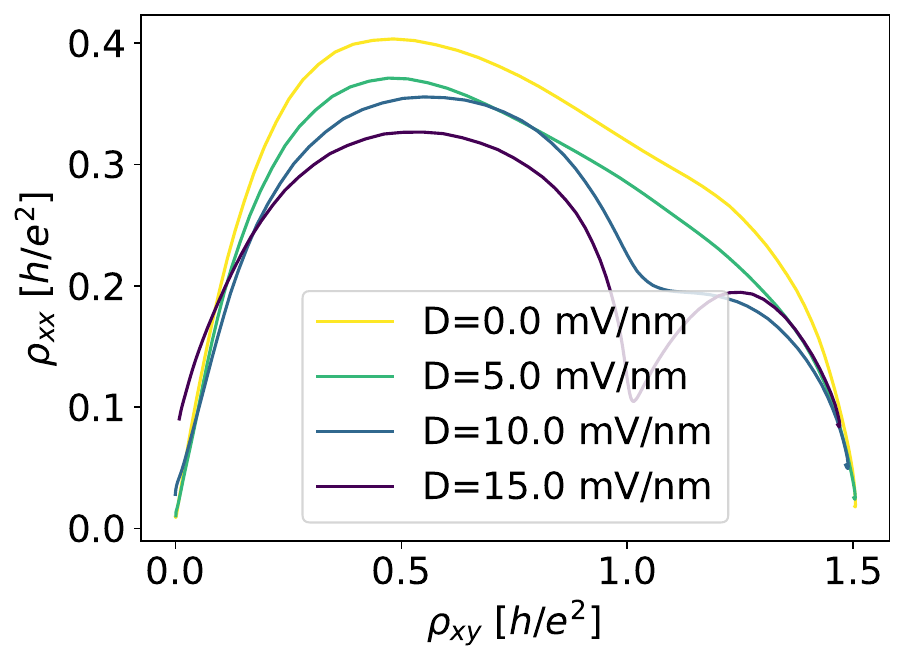}
    \caption{The faint feature near $\rho_{xy}=h/e^2$ in Fig.~\ref{fig:semicircle_thermal} is amplified by introducing a displacement field $D$. This stabilises the RIQHE phase, and the shape of the resistivity curve evolves towards the simulated behaviour in Fig.~\ref{fig:simulation_results} for a better-developed RIQHE phase [Data from Ref.~\cite{xu2026signaturesunconventionalsuperconductivitynear}].}
    \label{fig:finte_d_sweeps}
\end{figure}

In the zero-temperature limit of this model, $\rho_{xx}^\text{RI}\to0$, where we clearly see two distinct semicircles in Fig.~\ref{fig:simulation_results}(B), consistent with having two percolation transitions and scenario (B) for the low-temperature phase diagram proposed in Fig.~\ref{fig:phase_diagram_shapes}. While the experimental temperatures do not seem low enough for the RIQHE regions to develop sufficiently to clearly resolve the two transitions, the displacement field $D$ provides another tuning knob, with larger $|D|$ further stabilising the RIQHE. In Fig.~\ref{fig:finte_d_sweeps}, we plot the experimentally observed cut for a selection of values of $D$. 
When $|D|$ is increased, the RIQHE becomes more stable, and accordingly the shape of the curve in the $(\rho_{xy},\rho_{xx})$ plane moves towards the results obtained in Fig.~\ref{fig:simulation_results} when a well-developed RIQHE is present. We interpret the small kink still present in the $D=0$ sweeps of Fig.~\ref{fig:semicircle_thermal} as evidence consistent with a weak RIQHE signature present even at $D=0$.

\textit{Conclusions---}
{We have argued that the apparent FCI–SC transition in twisted MoTe$_2$ can be understood as an inhomogeneous three-phase problem involving the FCI, RIQHE, and SC. The evolution of the resistive peak with temperature and displacement field is naturally reproduced if the RIQHE has the lowest characteristic ordering energy scale and therefore becomes relevant only at lower temperatures. At sufficiently low temperatures, the RIQHE then forms an intervening phase between the FCI and SC down to zero displacement field. }

 We emphasize that our model contains relatively few arbitrary fit parameters. The value of $\rho_{xx}^\text{RI}$ is allowed to have an unspecified temperature dependence, but comparing the resulting family of curves to experiment can in principle be used to extract this microscopic dependence. The only independently chosen parameter is $w$. However, the family of curves generated by this model is quite robust to this choice of $w$, although its precise value affects the extracted $\rho_{xx}^\text{RI}(T)$.

The observed resistive peak remains broad, with an apparent width $\Delta\nu\simeq0.022$ as the temperature is lowered \cite{xu2026signaturesunconventionalsuperconductivitynear,senthil2026fractionalizedmetalsdopedanyons}, in contrast to the expectation for a single critical point in the absence of an extended intermediate phase. Within our picture, $\Delta\nu$ provides an estimate of the width of the intervening RIQHE phase. We therefore predict that, upon lowering the electronic temperature at $D=0$, the present single broad resistive peak $\rho_{xx}(\nu)$ should split into two narrow peaks separated by an RIQHE region, with vanishing longitudinal resistivity between them; in the $(\rho_{xy},\rho_{xx})$ plane, this corresponds to two semicircle-like transitions.

Finally, one feature of the data not explained by the model above is the linear slope in the $(\rho_{xy},\rho_{xx})$ plane near the FCI in Fig.~\ref{fig:semicircle_thermal}. Very close to the FCI, before disorder-induced phase mixing becomes important, the deviation from $\nu=-2/3$ can instead be viewed as a dilute gas of mobile anyons at density $\delta=|\nu+2/3|$. In a system with spontaneously broken time-reversal symmetry, such a gas can give rise to a resistivity $\rho_{xx}=a \delta,~\rho_{xy}=3h/2e^2-b\delta$, which explains the linear slope \cite{senthil2026fractionalizedmetalsdopedanyons}. The coefficients $a,b$ depend on microscopic details, and for some models of itinerant anyons, it is possible that $b=0$ \cite{fan_hidden_2026}. Thus, at very small filling deviations an anyon gas can contribute to a linear slope (if $b\neq0$), while at intermediate deviations of $\nu$ from $-2/3$ we see the puddle picture discussed above.

\begin{acknowledgments}
\textit{Acknowledgments---}
We are very grateful to Tingxin Li for sharing some of the raw data produced in Ref.~\cite{xu2026signaturesunconventionalsuperconductivitynear}. We thank Akshat Pandey for discussions. T. L. acknowledges funding from a Leverhulme Trust International Professorship grant (No. LIP-202-014) and
S. H. S. acknowledges support from EPSRC Grant No. EP/X030881/1.
\end{acknowledgments}

\bibliography{bib}

\section*{End Matter}

\appendix*

\section{Details of numerical calculations}
\label{app:simulation_implementation}

To generate the data in Fig.~\ref{fig:simulation_results}, we consider the electrostatic field on a cylinder of size $L_x\times L_y$. With $\phi(x,y)$ denoting the electrostatic potential, we impose $\phi(0,y)=V,~\phi(L_x,y)=0$, and $\phi(x,y+L_y)=\phi(x,y)$. We discretise this system into unit cells of size $a\times a$ ($a=0.2$ is found to be sufficiently small), and generate the quenched disorder as $V(\mathbf r)=\mu+\xi(\mathbf r)$, where $\xi(\mathbf r)$ is Gaussian with zero mean and covariance $\langle\xi(\mathbf r)\xi(\mathbf r')\rangle=\exp(-|\mathbf r-\mathbf r'|^2/2)$. 

Once the disorder field is generated, we apply Eq.~\ref{eq:phase_assignment_rule} to each of the $(L_x/a)\times(L_y/a)$ unit cells, creating a spatial profile of the resistivity tensor $\rho(\mathbf r)$. Defining $-\nabla\phi(\mathbf r)=\mathbf E(\mathbf r)=\rho(\mathbf r)\mathbf j(\mathbf r)$, we solve for $\phi$ under the aforementioned boundary conditions and $\nabla\cdot \mathbf j=0$, in a discretized finite-difference scheme. Although tools for solving this sparse linear system are readily available, we note that numerical stability is enhanced by introducing a small residual longitudinal resistivity to each phase, i.e. by letting $\rho_{xx}(\mathbf r)>0$ for all $\mathbf r$. We set the regulator to $\epsilon=0.025$, corresponding to a residual resistivity of $0.025h/e^2$, roughly matching the lowest values seen in experiment.

Given the resulting field, we compute $I_x=a\sum_y j_x(x,y)$ as the total longitudinal current (which is independent of $x$ by $\nabla\cdot \mathbf j=0$), and $I_y=a\sum_x j_y(x,y)$ as the total Hall current. The effective conductivity tensor is read off as $\sigma_{xx}=(L_xI_x)/(L_yV)$ and $\sigma_{xy}=I_y/V$. Each point in Fig.~\ref{fig:simulation_results} is obtained by averaging $\sigma_{xx}$ and $\sigma_{xy}$ over 200 independent disorder realisations with $L_x=100,~L_y=50$, and then inverting to obtain $\rho_{xx}=\langle \sigma_{xx}\rangle/(\langle \sigma_{xx}\rangle^2+\langle\sigma_{xy}\rangle^2)$ and $\rho_{xy}=-\langle \sigma_{xy}\rangle/(\langle \sigma_{xx}\rangle^2+\langle\sigma_{xy}\rangle^2)$, where $\langle\cdot\rangle$ denotes averaging over disorder realisations.

\end{document}